\documentclass[sn-aps,iicol,pdflatex]{sn-jnl}

\usepackage{graphicx}%
\usepackage{multirow}%
\usepackage{amsmath,amssymb,amsfonts}%
\usepackage{amsthm}%
\usepackage{mathrsfs}%
\usepackage[title]{appendix}%
\usepackage{xcolor}%
\usepackage{textcomp}%
\usepackage{manyfoot}%
\usepackage{booktabs}%
\usepackage{algorithm}%
\usepackage{algorithmicx}%
\usepackage{algpseudocode}%
\usepackage{listings}%
\usepackage{graphicx}%
\usepackage{multirow}%
\usepackage{amsmath,amssymb,amsfonts}%
\usepackage{amsthm}%
\usepackage{mathrsfs}%
\usepackage[title]{appendix}%
\usepackage{xcolor}%
\usepackage{textcomp}%
\usepackage{manyfoot}%
\usepackage{booktabs}%
\usepackage{algorithm}%
\usepackage{algorithmicx}%
\usepackage{algpseudocode}%
\usepackage{listings}%
\usepackage{bm,physics}
\usepackage{amsmath,amssymb}
\usepackage{ulem, color}
\usepackage{amsmath}
\usepackage{natbib}
\bmdefine{\ba}{a}
\bmdefine{\bb}{b}
\bmdefine{\bx}{x}
\bmdefine{\by}{y}
\bmdefine{\bz}{z}
\bmdefine{\bn}{n}
\bmdefine{\bp}{p}
\newcommand{\BM}{\begin{pmatrix}}
\newcommand{\EM}{\end{pmatrix}}

\begin{document}

\title {Existence of a fourth Airy elephant in  the nuclear rainbows for   $^{12}$C+$^{12}$C   scattering \\
}
\author*[1]{\fnm{S.} \sur{Ohkubo}}\email{ohkubo@rcnp.osaka-u.ac.jp}

\author[2]{\fnm{Y.} \sur{Hirabayashi}}

\affil*[1]{\orgdiv{Research Center for Nuclear Physics}, \orgname{Osaka University}, \orgaddress{\street{} \city{Ibaraki}, \postcode{567-0047}, \state{} \country{ Japan}}}

\affil[2]{\orgdiv{Information Initiative Center, Hokkaido University}, \orgname{Hokkaido University},\orgaddress{\street{} \city{Sapporo}, \postcode{060-0811 }, \state{} \country{Japan}}}

\pagenumbering{arabic}
\date{\today}
\pagenumbering{arabic}\pagenumbering{arabic}
\date{\today}
\pagenumbering{arabic}

\abstract{ 
The number of gross structures in the 90$^\circ$ excitation function for $^{12}$C+$^{12}$C elastic scattering—often called "Airy elephants"—has been of great interest. These
 structures are caused by refractive scattering and are separated by Airy minima. Their importance stems from their close relationship to the interaction potential between two $^{12}$C nuclei, which also describes the  molecular resonances of the compound system at lower energies.   Although a unique deep potential was usually  determined from
  rainbow scattering at higher energies, a puzzling discrepancy persisted: the energy at which the Airy minimum $A1$ crosses 90$^\circ$ was $E_{c.m.}\approx$67 MeV for $^{12}$C+$^{12}$C. This is remarkably low compared to approximately 100 MeV for both the
   $^{16}$O+$^{12}$C 
  and $^{16}$O+$^{16}$O systems. This question remained unanswered until the discovery of the secondary rainbow in the $^{12}$C+$^{12}$C system. We report for the first time that the highest energy at which the  dynamically generated Airy minimum of the secondary rainbow crosses 90$^\circ$ is about 100 MeV. This demonstrates that the fourth Airy elephant exists between the Airy minimum $A1$ of the primary nuclear rainbow and that, $A1^{(S)}$, of the secondary rainbow.
The long-standing problem concerning the Airy minima and Airy elephants has finally been resolved after decades of concern by recognizing the existence of a dynamically generated secondary rainbow in $^{12}$C+$^{12}$C scattering.
}




\maketitle
\section{Introduction}
\par
Gross structures observed in excitation functions provide important information about the structure of the nucleus. The famous structure observed by Bromley et al. \cite{Bromley1960} in $^{12}$C+$^{12}$C elastic scattering at $E_L$=6 - 29 MeV opened the field of molecular structure in heavy-ion systems. Quasi-molecular structures for the typical $^{12}$C+$^{12}$C  system, as well as $^{16}$O+$^{16}$O, were extensively studied \cite{Bromley1984,Betts1997}. The measurement of the 90$^\circ$ excitation function in $^{12}$C+$^{12}$C  scattering was extended to $E_L$=27 - 75 MeV by Reilly et al. \cite{Reilly1973}. Stokstad et al. \cite{Stokstad1975} further extended the measurement of angular distributions to $E_L$=75-130 MeV systematically and observed the development of gross structure in the 90$^\circ$ excitation function, which is not related to the resonances of the compound $^{12}$C+$^{12}$C system. Brandan et al. \cite{Brandan1990} were able to describe the evolution of the angular distributions with energy by using a global deep real potential. On the other hand, Morsad et al. \cite{Morsad1991}, measuring the excitation functions at $E_L$=60-120 MeV, found that the intermediate structure resonances disappear above $E_L$= 70 MeV, while the broad and irregular structure becomes a general feature of the interaction at higher energies.
%

\begin{figure}[t]
\centering
\includegraphics[keepaspectratio,width=8.0cm] {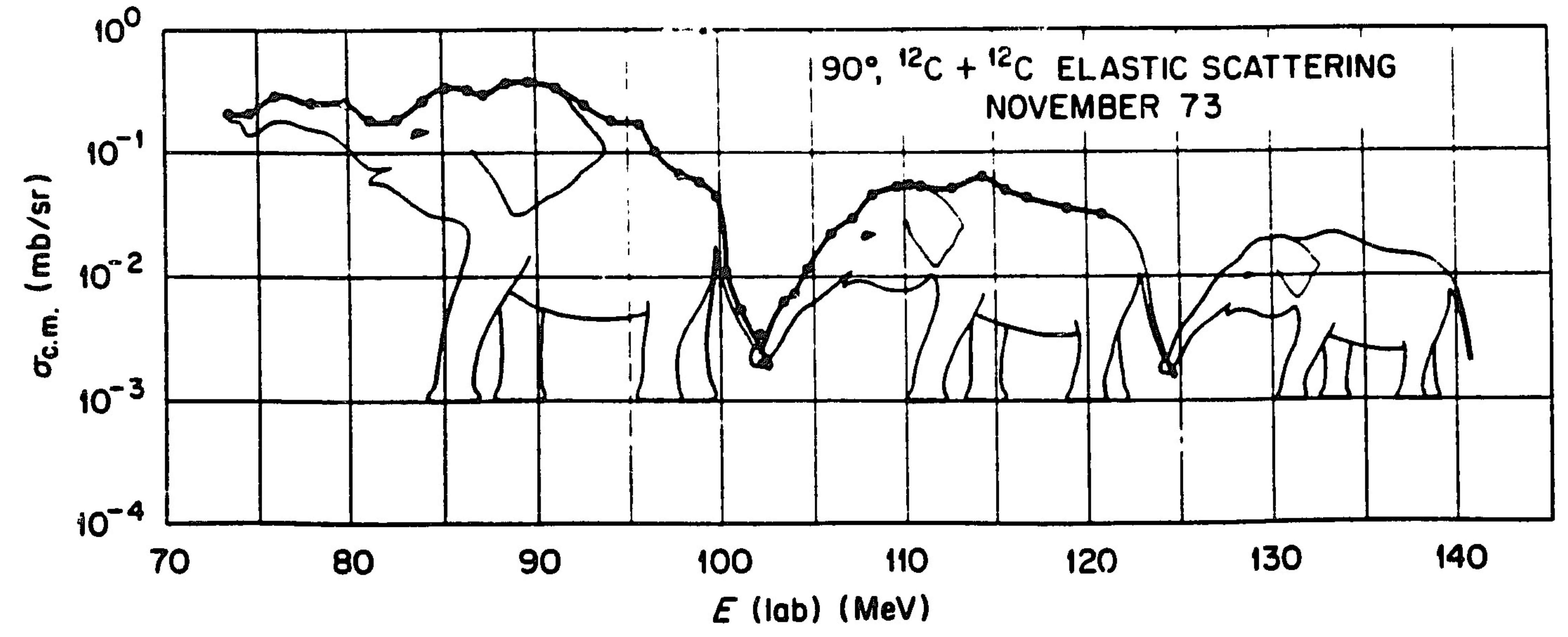}%
 \protect\caption{\label{fig1Elephant} {
 The familiar elephant interpretation \cite{Raman1974,McVoy1992} of the shape of the experimental 90$^\circ$ excitation function (points) in refractive $^{12}$C+$^{12}$C scattering. The line (elephant) is to guide the eye.
The minima separating the elephants correspond to the Airy minima.
 This figure is reproduced from Ref.~ \cite{McVoy1992} under STM permission from Elsevier publisher.
  }
  }
\end{figure}

  McVoy and Brandan \cite{McVoy1992}  analyzed the  angular distributions  of Ref.~\cite{Stokstad1975} and discussed the structure of the 90$^\circ$  excitation function from  viewpoint of  refractive scattering,  see Fig.~\ref{fig1Elephant},
    using a unique deep potential  in Ref.~\cite{Brandan1990} determined from the analysis of rainbow scattering above $E_L$=150 MeV and concluded that the most prominent minimum is
     unquestionably due to  Airy minimum $A2$ that crosses 90$^\circ$ at 102 MeV.  They expected a minimum due to $A3$ at near $E_L$=55  MeV.
They confidently predicted the existence of the Airy minimum  $A1$ at about $E_L$= 130 MeV as a final minimum.

Michel and Ohkubo \cite{Michel2004} systematically studied the energy evolution of the Airy minima in $^{12}$C+$^{12}$C scattering extensively at $E_L$=70-130 MeV by decomposing the farside scattering amplitude  into the internal wave and barrier wave components using the prescription in Refs.~\cite{Fuller1975, Brink1985,Brink1977,Albinski82} and showed that the so-called Airy minima in this energy region are, strictly speaking, the interference minima of the prerainbows proposed in Ref. \cite{Michel2002}. They also showed that the Airy minimum $A3$ crosses 90$^\circ$ at around $E_L$=75 MeV. Demyanova et al. \cite{Demyanova2010A,Demyanova2010B}, by measuring the rainbow scattering at 240 MeV up to large angles toward 90$^\circ$, concluded that there is no Airy minimum that crosses 90$^\circ$ above $E_L$= 130 MeV. 
Thus, the number of Airy minima that cross 90$^\circ$ seems to have been solved to be four ($A1$ - $A4$) decades after the humorous "Airy elephants".
 %

However, there remained a puzzle that  the   energy $E_{c.m.}\approx$67 MeV  that the $A1$ crosses  90$^\circ$ for $^{12}$C+$^{12}$C is  remarkably smaller compared with about $E_{c.m.}$=100 MeV for  $^{16}$O+$^{12}$C and 95 MeV for $^{16}$O+$^{16}$O. 
It may be  difficult to believe that the refractive scattering due to the mean-field potential changes so drastically among the neighbouring  three systems.
 Although since the first experimental observation  in Refs.~\cite{Brandan1982,Bohlen1982,Buenerd1982} tremendous studies \cite{Buenerd1982,Bohlen1982,Bohlen1985,Brandan1982,Brandan1988,Brandan1990,%
McVoy1992,Khoa1994,Brandan1997,Hassanain2008,Demyanova2010A,Demyanova2010B,%
Furumoto2012,Khoa2016,Hemmdan2021,Phuc2021} have been devoted to nuclear rainbow scattering for  $^{12}$C+$^{12}$C and the interaction potential that describes the rainbow scattering, this seeming puzzle has not been answered in the last decades.
Thus the number of the gross structures in the 90$^\circ$  excitation function  separated by the Airy minima, which evokes of the shape of an elephant and  have been called  Airy elephant humorously \cite{Raman1974,McVoy1992}, remained   as a puzzle concerning the  number of the Airy elephants.
 
\par
The purpose of this paper is to report, for the first time, the existence of an Airy elephant above  $E_L$= 130 MeV. We demonstrate this by showing that the highest energy at which the Airy minimum crosses 90$^\circ$ is  $E_L\approx$210 MeV ($E_{c.m.}$=105 MeV). This corresponds to the Airy minimum  $A1$ of the dynamically generated secondary bow, not the primary nuclear rainbow. This finding resolves the puzzle concerning the number of Airy minima in the  90$^\circ$  excitation function, the number of Airy elephants, and the previously reported, remarkably lower energy of the  $A1$ minimum crossing 90$^\circ$ for $^{12}$C + $^{12}$C compared with $^{16}$O+$^{12}$C and  $^{16}$O+$^{16}$O.

\section{The extended double folding model}
    
We use an extended double folding (EDF) model, which is given by 
\begin{eqnarray}
\lefteqn{V_{ij,kl}({\bf R}) =
\int \rho_{ij}^{\rm (^{12}C)} ({\bf r}_{1})\;
     \rho_{kl}^{\rm (^{12}C)} ({\bf r}_{2})} \nonumber\\
&& \times v_{\it NN} (E,\rho,{\bf r}_{1} + {\bf R} - {\bf r}_{2})\;
{\it d}{\bf r}_{1} {\it d}{\bf r}_{2} ,
\end{eqnarray}
\noindent where
$\rho_{ij}^{\rm (^{12}C)} ({\bf r})$ represents the diagonal ($i=j$) or transition ($i\neq j$)
 nucleon density of $^{12}$C which is calculated using the microscopic three $\alpha$ cluster model in the resonating group method \cite{Kamimura1981}. 
 In the coupled channel (CC)  calculations, we take into account the excitation of the  $2^+$ (4.44 MeV), $3^-$ (9.64 MeV) and $4^+$ (14.08 MeV) states of $^{12}$C. For the effective nucleon-nucleon interaction ($v_{\rm NN}$), we use the DDM3Y-FR interaction \cite{Kobos1982,Kobos1984,Bertsch1977}, which accounts for the finite-range nucleon exchange effect \cite{Khoa1994}. We introduce the normalization factor $N_R$ 
  \cite{Satchler1979,Brandan1997} for the real double folding potential. 
  The Coulomb folding potential $V^C_{ij,kl}(${\b R}$)$ is similarly calculated in the double folding model of Eq.~(1) by replacing $v_{NN}$ by  $v_{Coul}$.
  An imaginary potential with a Woods-Saxon volume-type (nondeformed) form factor is phenomenologically introduced to account for the effect of absorption due to other channels.

 \section{Airy minimum at 90$^\circ$ due to secondary bow }
    
\begin{figure}[t]
\centering
\includegraphics[keepaspectratio,width=8.0cm] {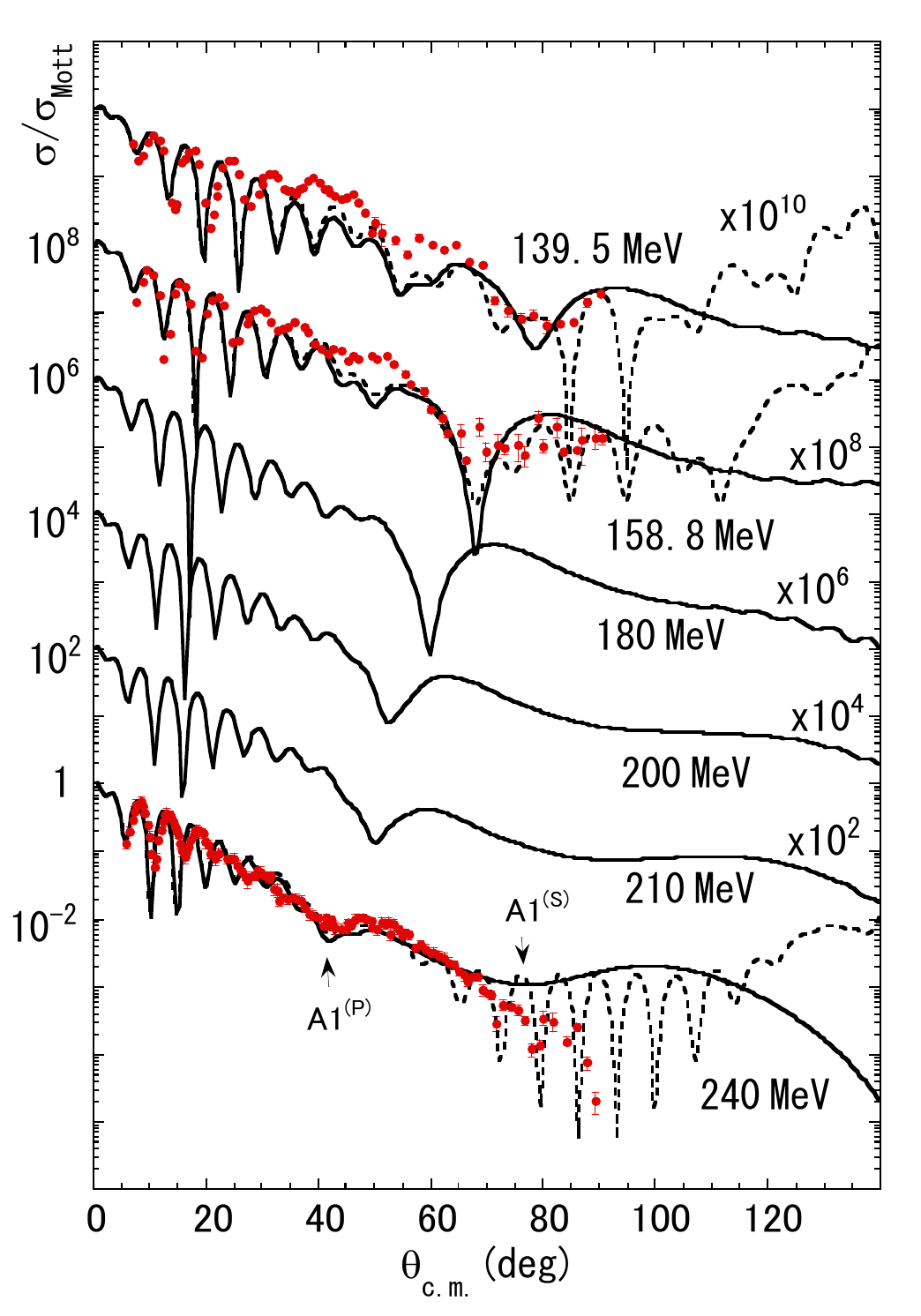}%
\protect\caption{\label{fig2}  {
Angular distributions calculated with (dashed lines) and without (solid lines) symmetrization, using the CC method with EDF, are displayed for energies above 139.5 MeV. $A1^{(P)}$  and $A1^{(S)}$  denote the Airy minimum of the primary nuclear rainbow and the secondary bow, respectively. Notably, at 210 MeV, the Airy minimum of the secondary bow appears at 90$^\circ$. These calculations are compared with experimental data (points) at 139.5 MeV and 158.8 MeV \cite{Kubono1983,Kubono1985}, and at 240 MeV \cite{Bohlen1985,Demyanova2010A,Demyanova2010B}. 
The unsymmetrized angular distributions (solid lines) are presented as a ratio to Rutherford scattering (vertical axis $\sigma/\sigma_R$).
  }
  }
\end{figure}

\par
The nuclear rainbow, a Newton's zero-order rainbow \cite{Michel2002}, is caused solely by refraction due to the nuclear potential. This differs from a meteorological rainbow \cite{Descartes, Newton, Airy1838, Nussenzveig1977,Adam2002} formed in a droplet, which involves reflections. Although nuclear rainbow scattering has been extensively studied \cite{Khoa2007}, proving powerful in unambiguously revealing the internuclear potential up to the internal region and greatly aiding in understanding cluster structure \cite{Michel1998, Ohkubo1999}, in principle, no secondary rainbow is expected in the standard theory of nuclear rainbow scattering \cite{Ford1959, Newton1966}. This theory predicts deflection functions with only one extremum, corresponding to Newton's zero-order rainbow. In fact, since the first discovery of the nuclear rainbow \cite{Goldberg1974}, no one could have imagined the existence of a dynamically generated nuclear rainbow, especially when the passage of the Airy minimum in the 90$^\circ$ excitation function and the number of Airy elephants have been discussed.

\par
The existence of the Airy minimum of the dynamically created secondary bow was first discovered by Ohkubo and Hirabayashi in $^{16}$O+$^{12}$C rainbow scattering \cite{Ohkubo2014}. They achieved this by investigating an anomalous bump in the fall-off region, on the dark side of the primary nuclear rainbow, which solved the puzzle surrounding the family of deep potentials that reproduce the experimental angular distributions. The Airy minimum attributed to the secondary bow was further confirmed in  $^{13}$C + $^{12}$C rainbow scattering  \cite{Ohkubo2015B}.
The third discovery of the Airy minimum from the secondary bow in $^{12}$C + $^{12}$C rainbow scattering  \cite{Ohkubo2025} definitively confirmed this concept. In this case, the bright side of the bow breaks up into ripples due to symmetrization of the two bosonic identical nuclei, which hindered the recognition of the secondary bow in the observed angular distributions.

\par
Figure~\ref{fig2} displays the angular distributions for $^{12}$C + $^{12}$C scattering, calculated using the EDF in the coupled-channel method. This calculation includes six-channel couplings of 
 ($^{12}$C(${ I}^\pi$), $^{12}$C(${ J}^\pi$)) = 
  ($0^+$,  $0^+$), ($0^+$,  $2^+$),  ($0^+$,  $3^-$), ($0^+$,  $4^+$), ($2^+$,  $2^+$), and ($2^+$ ,  $4^+$). 
Mutual excitations of the  $2^+$ state, as well as the orientation effect, are included.  
The parameters employed in the calculations were interpolated between $N_R$=1.05  and  $W$=17 at $E_L$=139.5 MeV, and $N_R$=1.13 and $W$=18 at $E_L$=240 MeV.
 A radius parameter $R_W$=5.6 fm and a diffuseness parameter $a_W$=0.7 fm for the imaginary potential are used. 
The position of the Airy minimum is insensitive to the imaginary potential; therefore, no parameter search was performed on the imaginary potential to improve fits to the experimental data at individual energies.
The angular distributions in this energy region are  due to farside refractive scattering at large angles \cite{Ohkubo2025}. For example, the fall-off of the angular distribution beyond 
   $\theta_{\rm c.m.}$=50$^\circ$ is the dark side of the primary nuclear rainbow.
The calculation reproduces the experimental angular distribution at 240 MeV, where the first-order Airy minimum  $A1^{(P)}$ ($A1$) of the primary nuclear rainbow is determined at 43$^\circ$ and the Airy minimum of the secondary bow  $A1^{(S)}$ is located at $\theta_{\rm c.m.}$=77$^\circ$. 
We note that the depth of the Airy minimum $A1^{(P)}$  is affected by symmetrization in the lower energy region. This improves the fits to the experimental data compared to the unsymmetrized results (solid lines). Specifically, at 158.8 MeV, the minimum becomes shallower, and at 139.5 MeV, it's obscured.
 Due to the symmetrization of the two bosonic identical nuclei, the bright bump of the secondary bow is broken up into symmetrization ripples. The existence of  $A1^{(S)}$ can be seen in the angular distributions calculated without symmetrization (solid lines).
As the incident energy increases, the positions of  $A1^{(P)}$ of the primary rainbow and $A1^{(S)}$ of the secondary rainbow move forward. By investigating the energy evolution of the Airy minima, as seen in Fig.~\ref{fig2}, one finds that $A1^{(S)}$ passes 90$^\circ$ at   $E_L\approx$210 MeV.

 \section{Evolution of the Airy minima and existence of the fourth Airy elephant}
 \par
\begin{figure}[t!]
\centering
\includegraphics[keepaspectratio,width=7.6cm] {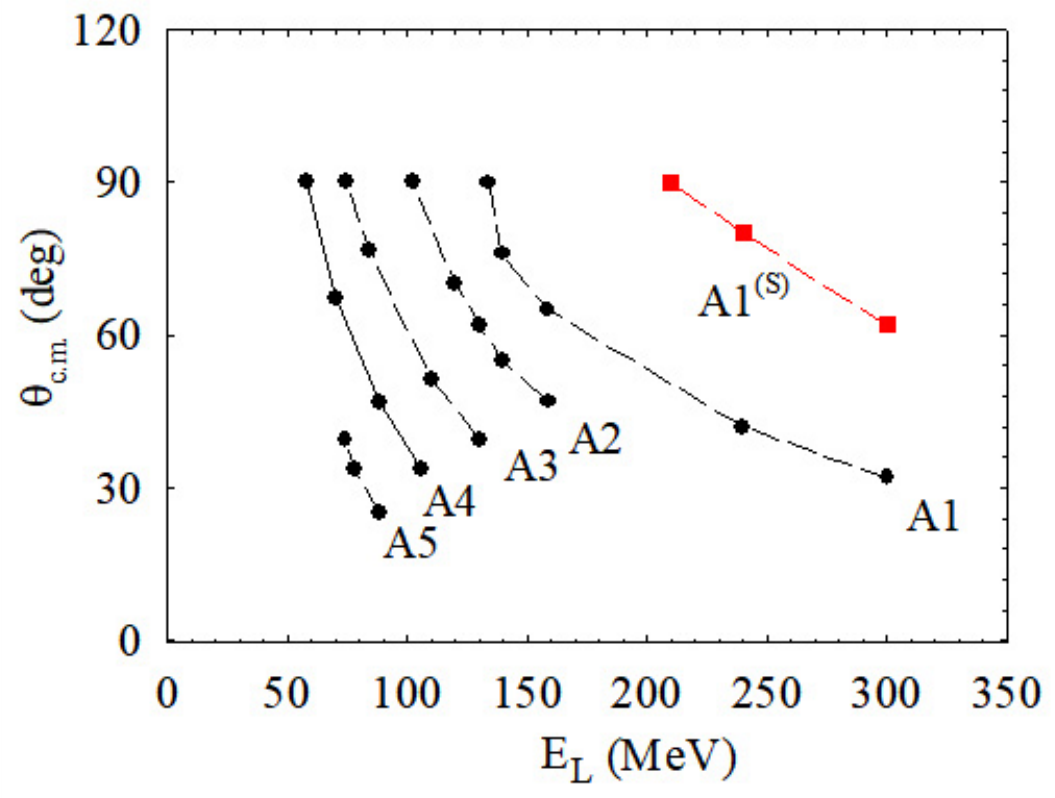}%
 \protect\caption{\label{fig3Airy-Edep} {
The energy evolution of the Airy minima for $^{12}$C + $^{12}$C elastic scattering is shown. The Airy minima for $E_L$ $<$130 MeV are taken from Ref.~\cite{Michel2004}, which reproduces the energy evolution of the experimental angular distributions. The Airy minima for $A1$ and $A2$  at $E_L>$130 MeV are derived from experimental data presented in Refs.~\cite{Bohlen1982,Kubono1983,Kubono1985,Bohlen1985,Demyanova2010A,Demyanova2010B}.
The labels $A1$, $A2$, $A3$,  etc., denote the order of the Airy minima. $A1^{(S)}$ of the secondary bow at $E_L$=210, 240, and 300 MeV are determined by the six-channel CC calculations \cite{Ohkubo2025}. The line is to guide the eye.
  }
}
\end{figure}

 Fig.~\ref{fig3Airy-Edep} systematically displays the energy evolution of the Airy minima $A1$-$A5$ of the primary nuclear rainbow and $A1^{(S)}$ of the secondary bow. The highest order Airy minimum, $A5$, observed in the experiment disappears at lower energies and does not pass through 90$^\circ$. Consequently, the highest Airy minimum that appears at 90$^\circ$ is $A4$ at $E_L\approx$58 MeV.
Thus, four gross structures, or "Airy elephants," appear in the 90$^\circ$ excitation function for $^{12}$C + $^{12}$C elastic scattering. The emergence of the fourth Airy elephant between  $A1$ (at around $E_L\approx$130 MeV) and $A1^{(S)}$ at around 210 MeV is solely due to the secondary bow, which would not have been expected without the study of a secondary bow in nuclear rainbow scattering.

\par
The present results provide the definitive answer to the Airy elephant interpretation of the 90$^\circ$ excitation function in $^{12}$C + $^{12}$C scattering, a topic discussed over the past few decades. Although McVoy and Brandan \cite{McVoy1992} concluded from a systematic analysis at 70 MeV $<E_L < $130 MeV, using a unique deep mean-field potential, that the elephant-shaped gross structure lies between the Airy minimum  $A2$ at $E_L$=102 MeV and $A1$ at $E_L$=130 MeV, it was later identified as the third Airy elephant. This is because the minimum at 74 MeV, which they considered a diffraction minimum and not an Airy minimum (see Fig.~7 in Ref.~\cite{McVoy1992}), was subsequently found to be an Airy minimum $A3$ in a systematic analysis by Michel and Ohkubo \cite{Michel2004}.
Their conjecture \cite{McVoy1992} that "the final smaller Airy elephant will exist above 130 MeV with a hump at 140 MeV," famously stating, "Woolly mammoths may be extinct but Airy elephants are alive and well", has been refuted by Demyanova et al. \cite{Demyanova2010A,Demyanova2010B}. Their study of the Airy minima in rainbow scattering at 240 MeV led them to conclude that no Airy minimum crosses 90$^\circ$ above $E_L$= 130 MeV.

 \par 
 
\begin{figure}[t!]
\centering
\includegraphics[keepaspectratio,width=7.8cm] {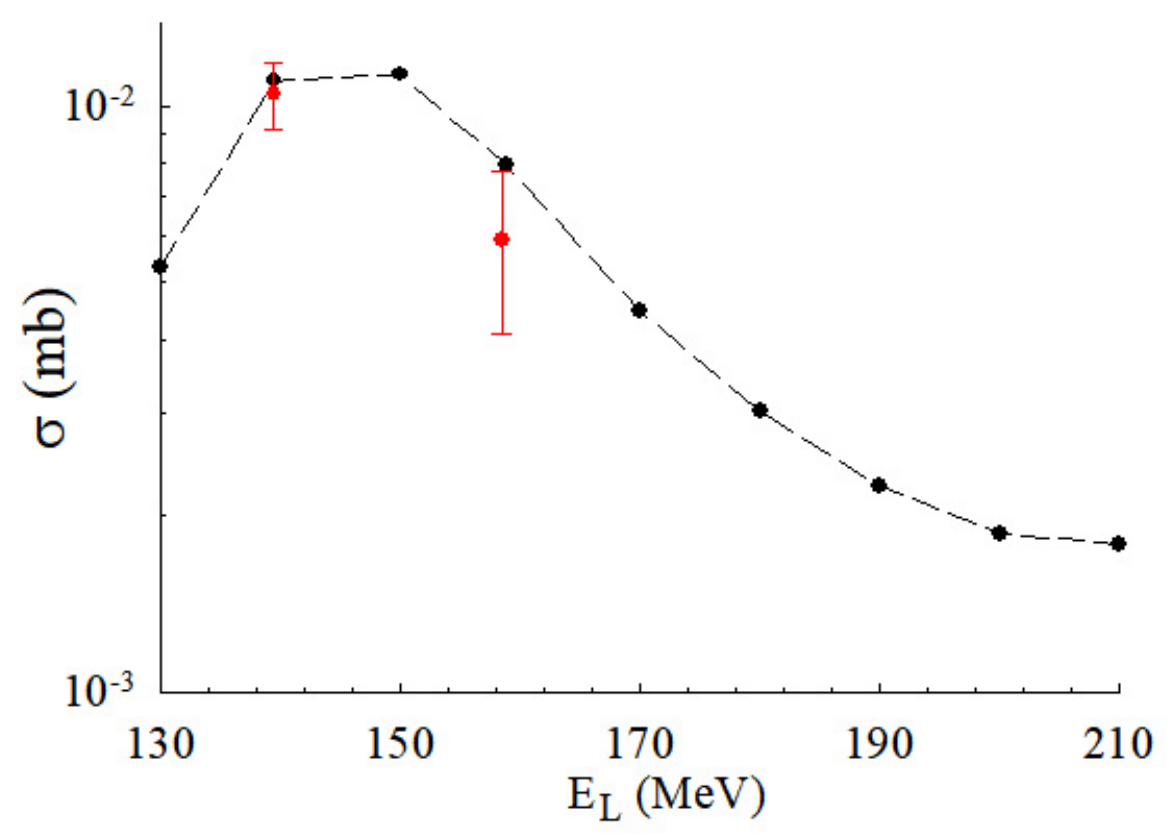}%
 \protect\caption{\label{fig4Elephant} {
The 90$^\circ$ excitation function for $^{12}$C + $^{12}$C scattering, calculated with the six-channel CC method using the EDF, exhibits a gross structure corresponding to the fourth Airy elephant. The line serves as a guide to the eye.
 Experimental data, including error bars, at 139.5 MeV and 158.8 MeV are derived from Refs. \cite{Kubono1983,Kubono1985}.
   }
}
\end{figure} 

\par
We display in Fig.~\ref{fig4Elephant} the excitation function at  90$^\circ$, showing the gross structure associated with the fourth Airy elephant, calculated using the EDF in the CC method. In the calculations the parameters were interpolated in the range of $E_L$=139.5 MeV to 240 MeV, and extrapolated for $E_L$ values below 139.5 MeV, as detailed in section 3.
 The maximum of the cross section, associated with the Airy maximum   of the primary nuclear rainbow, is located between $E_L$=140 and 150 MeV.
The first Airy elephant lies between $A4$  (at around $E_L\approx$58 MeV) and $A3$ (at around $E_L\approx$75 MeV), which was not identified in Ref.~\cite{McVoy1992}. The second Airy elephant is located between $A3$ and $A2$, and the third Airy elephant is also between $A2$ and $A1$.
It is now clear that the fourth Airy elephant exists for the $^{12}$C + $^{12}$C system. It is highly desirable to experimentally measure the excitation function at 90$^\circ$ in $^{12}$C + $^{12}$C scattering between $E_L$=130 - 220 MeV. After long and confusing discussions, the problem of the Airy elephant has now been finally resolved.

\section{Evolution of the Airy minima in  $^{12}$C+$^{12}$C, $^{16}$O+$^{12}$C, and $^{16}$O+$^{16}$O} 

\begin{figure}[t!]
\centering
\includegraphics[keepaspectratio,width=7.8cm] {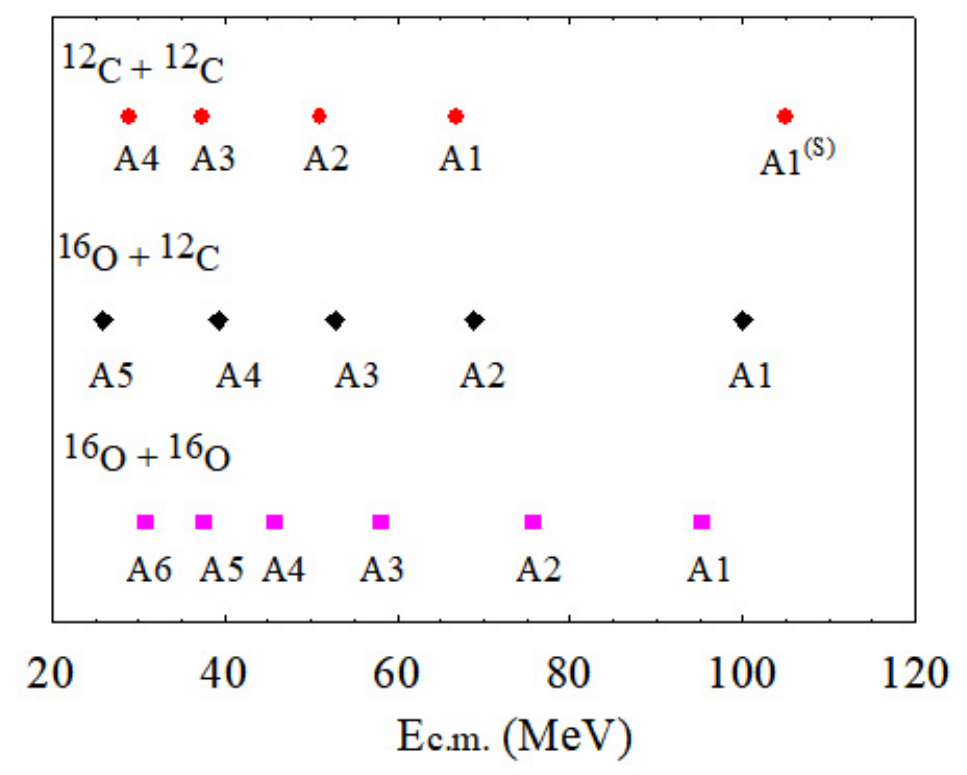}%
 \protect\caption{\label{fig5-12C16OAiry} {
 The Airy minima of the  primary nuclear rainbows in the experimental 90$^\circ$ excitation functions for elastic scattering in the $^{12}$C+$^{12}$C, $^{16}$O+$^{12}$C, and $^{16}$O+$^{16}$O systems,  where the labels  $A1$, $A2$, $A3$, { etc.} denote the order of the Airy minima of the primary nuclear rainbows, and the Airy minimum of the secondary rainbow $A1^{(S)}$ for the $^{12}$C+$^{12}$C system   calculated in the CC  method. Extended from Fig.~1 of Ref. \cite{Demyanova2010B}.
 }
}
\end{figure}
\par 
In Fig.~\ref{fig5-12C16OAiry}, the energies at which the Airy minima appear at 90$^\circ$ are compared for the three systems: $^{12}$C+$^{12}$C, $^{16}$O+$^{12}$C, and $^{16}$O+$^{16}$O. The highest order of the Airy minima is $A4$ for  $^{12}$C+$^{12}$C, $A5$ for $^{16}$O+$^{12}$C, and $A6$ for $^{16}$O+$^{16}$O.
Although the highest energy $E_{c.m.}\approx$67 MeV for $A1$ for $^{12}$C+$^{12}$C is remarkably smaller than the corresponding energies for  $^{16}$O+$^{12}$C ($E_{c.m.}$=100 MeV) and $^{16}$O+$^{16}$O ($E_{c.m.}$=95 MeV), Demyanova { et al.}~\cite{Demyanova2010A,Demyanova2010B} concluded that the last Airy extremum (referred to as the fourth Airy elephant in their notation) does not appear in the higher energy region above  $E_L$=130 MeV, which is the last extremum shown in Fig.~1.
Their conclusion is correct concerning the  $A1$ of the primary nuclear rainbows due to the static mean-field potential. However, because the secondary rainbow is generated dynamically, $A1^{(S)}$ of the secondary rainbow at $E_L\approx$210 MeV is the last Airy minimum that crosses 90$^\circ$.
Nature seems to prefer the last Airy minimum around $E_{c.m.}$=100 MeV for these three neighboring systems ($^{12}$C+$^{12}$C, $^{16}$O+$^{12}$C, and $^{16}$O+$^{16}$O). This might not be accidental since $A1^{(S)}$ of the secondary rainbow in $^{16}$O+$^{12}$C rainbow scattering, which has been observed at angles smaller than 90$^\circ$ \cite{Ohkubo2014}, does not cross 90$^\circ$.

 \section{Summary}
 For a long time, the number of gross structures—often referred to as "Airy elephants"—separated by the Airy minima in the 90$^\circ$ excitation function for $^{12}$C+$^{12}$C refractive scattering has been a persistent concern. This remained true even after a deep $^{12}$C+$^{12}$C potential was established from the analysis of rainbow scattering at high energies. Furthermore, the energy $E_{c.m.}\approx$67 MeV, at which the Airy minimum $A1$ crosses 90$^\circ$ for $^{12}$C+$^{12}$C, has been a puzzle due to being  remarkably lower compared to approximately 100 MeV for both $^{16}$O+$^{12}$C  and  $^{16}$O+$^{16}$O. 
 
  We systematically studied the evolution of the  Airy minima of both the primary nuclear rainbow and the secondary rainbow. Using an extended double folding model within the coupled channel method, it was shown that the Airy minimum of the secondary rainbow crosses 90$^\circ$ at  $E_L\approx$210 MeV. We found that the 90$^\circ$ excitation function, consistent with experimental data, exhibits a gross structure between the  Airy minimum $A1$ of the primary nuclear rainbow and the Airy minimum  $A1^{(S)}$ of the secondary rainbow, which corresponds to the fourth Airy "elephant."
  
The long-standing problem concerning the Airy minima and Airy elephants has finally been resolved after decades of concern by recognizing the existence of a dynamically generated secondary rainbow in $^{12}$C+$^{12}$C scattering.

 \begin{itemize}
\item Funding
No funding.
\item Conflict of interest/Competing interests 
Not applicable
\item Consent for publication
The authors agree with publication of the manuscript in the traditional publishing model.
\item Data availability Statement
This manuscript has no associated data or the data will not be deposited. [Authors’ comment: This is a theoretical study using  published data, and all data information is properly referenced.]
\end{itemize}

\end{document}